\documentclass[10pt,twocolumn]{z-article} %

\usepackage{S/oneinchmargins}
\usepackage{times}
\usepackage{relsize}
\usepackage{enumerate}
\usepackage{graphicx}
\usepackage{url}
\usepackage[usenames,dvipsnames]{xcolor}
\usepackage[bookmarks=false]{hyperref}

\hypersetup{colorlinks=true,
citecolor=Maroon,
linkcolor=Green,
urlcolor=Maroon}

\usepackage{breakurl}
\usepackage{setspace}
\usepackage{rotating}

\usepackage{floatflt}
\usepackage{wrapfig}
\usepackage{alltt}
\usepackage{epstopdf}
\usepackage{subfigure}

\usepackage{fancyvrb}
\usepackage{caption}
\usepackage{stfloats}
\usepackage{booktabs}
\usepackage{graphicx}

\usepackage{makecell}

\usepackage{etoolbox}
\apptocmd{\thebibliography}{\raggedright}{}{}

\begin{document}

\newcommand{\infokernel}{\mbox{infokernel}}
\newcommand{\unix}{{\sc Unix}}
\newcommand{\dare}{DARE}
\def \fate {\textsc{Fate}}
\def \late {\textsc{Late}}
\def \lights {\textsc{LigHTS}}
\def \samsup {Sam$^{Zk}$}
\def \samsub {Sam$_{Zk}$}

\newcommand{\vtwenty}{\vspace{20pt}}
\newcommand{\vfifteen}{\vspace{15pt}}
\newcommand{\vten}{\vspace{10pt}}
\newcommand{\vfive}{\vspace{5pt}}
\newcommand{\vthree}{\vspace{3pt}}

\newcommand{\vminfive}{\vspace{-5pt}}
\newcommand{\vminten}{\vspace{-10pt}}
\newcommand{\vminfifteen}{\vspace{-15pt}}
\newcommand{\vmintwenty}{\vspace{-20pt}}

\newcommand{\ub}[1]{\underline{{\bf #1}}}
\newcommand{\ts}[1]{{\tt{\small#1}}}
\newcommand{\bquote}{\vspace{-0.25cm} \begin{quote}}
\newcommand{\equote}{\end{quote}\vspace{-0.2cm} }
\def \sec {\S}
\def \yes {$\surd$}

\def \nospace {
  \setlength{\itemsep}{0pt}
  \setlength{\parskip}{0pt}
  \setlength{\parsep}{0pt}
}

\newcommand{\mypara}[1]{\vspace{0.05cm}\noindent{\bf {#1}:}~}
\newcommand{\supsection}[1]{\noindent{\Large{\bf #1}}\vten}

\newenvironment{enumerate2}{
  \begin{enumerate}
  \setlength{\itemsep}{1pt}
  \setlength{\parskip}{0pt}
  \setlength{\parsep}{0pt}
}{
  \end{enumerate}
}

\newenvironment{itemize2}{
  \begin{itemize}
 \renewcommand{\labelitemi}{-}
  \setlength{\itemsep}{1pt}
  \setlength{\parskip}{0pt}
  \setlength{\parsep}{0pt}
}{
  \end{itemize}
}


\newcommand{\notes}[1]{\textcolor{darkgray}{{\small {\em (Notes: #1)}}}}
\newcommand{\hsg}[1]{\textcolor{red}{{\small {\bf (HSG: #1)}}}}
\newcommand{\thanh}[1]{\textcolor{red}{{\small {\bf (THANH: #1)}}}}
\newcommand{\todo}[1]{\textcolor{red}{{\small {\bf (TODO: #1)}}}}
\newcommand{\newtext}[1]{\textcolor{red}{#1}}
\newcommand{\bluetext}[1]{\textcolor{blue}{#1}}

\def \vvvnb {\vfifteen \noindent $\bullet$~}
\def \vvnb {\vten \noindent $\bullet$~}
\def \vnb {\vfive \noindent $\bullet$~}

\def \mb {\vspace{8pt}\nb}
\def \tb {\vspace{8pt}\nb}

\def \vvn {\vten \noindent}
\def \vn {\vfive \noindent}
\def \nb {\noindent $\bullet$~}
\def \ni {\noindent}

\newcommand{\hypo}[1]{
\begin{quote}
\stepcounter{HYPO}{\bf Hypothesis \arabic{HYPO}:} 
{\em #1}
\end{quote}
}

\newcommand{\taskformat}[2]{#1\textsc{#2}}

\newcommand{\task}[3]{
\begin{quote}
\phantomsection
\hypertarget{task#1#2}{}
{\bf Task \taskformat{#1}{#2}:} 
{\em #3}
\end{quote}
}

\newcommand{\tasklink}[2]{\hyperlink{task#1#2}{\taskformat{#1}{#2}}}

\newcounter{HYPO}
\newcounter{TASK}

\newcommand{\rs}{{ResearchStaff$_1$}}
\newcommand{\pd}{{\bf Postdoc$_1$}}
\newcommand{\raOne}{{\bf RA$_1$}}
\newcommand{\raTwo}{{\bf RA$_2$}}
\newcommand{\ndv}{{\bf NDV}}
\newcommand{\ug}{{\bf Undergrad$_1$}}


\newcommand{\sssubsection}[1]{\vten\textbf{\large{\textsc{#1}}}}

\newcommand{\emptypage}{
\newpage
\thispagestyle{empty}
(empty page)
}

\newcommand{\myrotate}[1]{\begin{rotate}{90} {\bf #1} \end{rotate}}

\newcommand{\mycaption}[3]{
\caption{
\label{#1}
{\bf #2. } 
{\em \small #3}
}}

\newcommand{\eg}{\textit{e.g.}}
\newcommand{\ie}{\textit{i.e.}}
\newcommand{\etal}{\textit{et al.}}
\newcommand{\etc}{etc.}

\newcommand{\symstar}{$^{\star}$}
\newcommand{\symtwostars}{$^{**}$}
\newcommand{\symthreestars}{$^{***}$}
\newcommand{\symdag}{$^{\dag}$}
\newcommand{\symddag}{$^{\ddag}$}
\newcommand{\symmath}{$^{\mathsection}$}
\newcommand{\tool}{\textsc{StorRep}}
\newcommand{\flashnet}{\textsc{FlashNet}}

\newcounter{Xcounter}
\newcommand{\xxxreset}{\setcounter{Xcounter}{1}}
\newcommand{\xxx}{
\textcolor{red}{
\textbf{XXX$_{\arabic{Xcounter}}$}\stepcounter{Xcounter}}}


\def \mytitle {\tool{}: Storage Research Experiment Patterns on Chameleon Cloud and Trovi}

\title{\textsf{\textbf{\mytitle}}}



\author{Ray A. O. Sinurat\symdag\footnote{The authors contribute equal amount of work and are sorted alphabetically based on their first names.}, Yuyang Huang\symdag$^{*}$, Nanqinqin Li$^{\star}$, Mark Powers\symmath\symdag, Michael Sherman\symmath\symdag,\\Kate Keahey\symmath\symdag, Haryadi S. Gunawi\symdag }

\date{
\begin{tabular}{ccc}
\symdag University of Chicago & 
$^{\star}$Princeton Unversity &
\symmath Argonne National Laboratory
\end{tabular}
}

\maketitle

\begin{abstract}

{\em Storage experiments are vital to advancing storage research, but creating extensible and reproducible storage artifacts can be a challenging task. Our research has shown that only 1\% of SSD simulator-based experiences are packaged and 0.5\% of them can be easily reproduced. The lack of such artifacts without proper reproducibility can significantly impede the advancement of storage research. The biggest challenges in these types of experiments are ensuring that we have the correct environment to conduct them and creating extensible experiments that can be built upon in future research. To address this issue, we introduce \tool{}, a thorough study that provides six extensible and reproducible storage experiment artifacts that serve as the foundation for further storage research, utilizing the Chameleon infrastructure. Our study offers experiment patterns and guidelines that can help researchers create transparent and dependable storage experiments. We have successfully integrated our methods in several experiments in multiple community and educational events over several years and produced publicly accessible artifacts that can be extended and fully reproduced without any restrictions.}

\end{abstract}

\vten

\section{Introduction}
In the computer science community, the evaluation of research projects typically involves the use of \textit{artifacts}, which consist of a combination of environment configuration (e.g. hardware and software setup and configuration), experiment scripts, and analysis of the experiment output (e.g. generation of figures).
\textit{Reproducing} the artifacts described in the research papers plays a crucial role in helping other researchers understanding and evaluating the research projects. However, such process can be difficult since the researchers may not be able to match the exact environment configuration with the artifacts' designed configuration and such discrepancy in the environment configurations can lead to results inconsistent with the original papers' claimed results, or completely incorrect results (\ie{} lead to the wrong/opposite conclusion). 
This significantly hampers the communication and mutual understanding within our community.

In the recent decade, the situation has drawn many researchers' attention. Moreover, there have been many efforts in improving the \textit{reproducibility} for papers' artifact from both the author-side (\eg{} developing an minimal-effort platform to package the artifact \cite{Chameleon-Trovi, Chameleon-PythonCHI}) and reader-side (\eg{} serving open testbeds that provide numerous types of hardware to help the research match the environment configuration\cite{Chameleon-Cloud, Duplyakin+2019-CloudLab}).
Moreover, an increasing number of conferences are recommending and assigning "Artifact Evaluation" or similar badge, which further highlights the importance of reproducibility in our community.

Despite increasing attention and interest in the computer science community, reproducibility in the storage research field has been progressing at a relatively slow pace. We conducted an case study on the research papers' artifacts related to Solid State Drives (SSDs) that were published over the last decade. We focused on 195 artifacts that claimed to run experiments on SSD simulators, such as SSDsim \cite{Hu+2011-SSDsim}, SimpleSSD \cite{Jung+2017-Simplessd, Gouk+18-Amber}, and NANDFlash \cite{Jung+2012-NANDFlashsim} since these simulators can be run on almost any device equipped with CPU and memory, and, therefore, the artifacts that use them should be easier to reproduce compared to other types of artifacts that require special hardware and software configurations.

To evaluate the reproducibility of these artifacts, we categorized them based on 6 criteria: \textit{Public}, \textit{Compilation}, \textit{Execution}, \textit{Automation}, \textit{Result}, and \textit{Package}. Surprisingly, we found that out of the 195 artifacts, only one (0.5\% of the total artifacts) satisfied all 6 criteria and was deemed reproducible. We discuss our investigation and its implications further in Section~\ref{sec:ssd-bleak-state}.

We understand the importance of conducting experiments that are not only accurate and insightful but also reproducible. As researchers, it is our responsibility to set up the ideal environment for experimentation, create experiment scripts, develop analysis and evaluation tools, and provide expected outcomes from these experiments (\ie{} artifacts). However, we all know that this can be a rigorous and challenging process. Fortunately, Chameleon \cite{Chameleon-Cloud} platform addresses these issues by integrating reliable and efficient tools that help researchers accomplish these arduous tasks with ease. 

One of the most challenging aspects of reproducing artifacts is providing the desired environment. Chameleon helps researchers overcome this hurdle by offering multiple types of storage and various types of nodes such as bare-metal machines and virtual machines, among other configurations. Additionally, Chameleon embraces the OpenStack \cite{OpenStack} ecosystem, which provides a better user experience and ease of managing environments. This means that researchers can now focus on their experiments without worrying about environment configurations.

One such feature of Chameleon is the integration of the Jupyter Notebook Interface \cite{JupyterInterface}, a widely recognized platform for running data science tasks. This integration allows researchers to run Python codes or shell scripts that can aid in the packaging of their experiments seamlessly. Moreover, Chameleon also offers the ability for users to share their experiment artifacts publicly on the Chameleon Trovi \cite{Chameleon-Trovi} platform, providing other users with the opportunity to reproduce their artifacts within the Chameleon infrastructure. This level of collaboration and transparency creates a vibrant and engaging community of researchers working together to achieve common goals.


Our research endeavors have led us to leverage the Chameleon platform for conducting and packaging our artifacts. By studying the four basic patterns provided by Chameleon in their blog \cite{Chameleon-BasicPatterns}, we were able to take advantage of the platform's infrastructure capabilities, including the straightforward setup of experiment containers and the efficient management of spawned server instances. Ultimately, we present \textbf{\tool{}} --- six storage-related artifacts that can be quickly and easily shared, reproduced and extended within the platform.

\tool{} includes (1) \textit{Simple Flexible I/O (FIO) \cite{Axboe-FIO} Benchmark}: an artifact to run FIO benchmark on RAM disk (mounted with \textit{tmpfs}) and a single SSD device, (2) \textit{Simple File System Benchmark}: an artifact to run multiple FileBench \cite{Filebench-FileBench} workloads on ext4 file system, (3) \textit{FIO Benchmark}: an artifact extended the Simple FIO Benchmark to run FIO Benchmark on a RAID-0 SSDs array, (4) \textit{File Systems Benchmark}: an artifact to simulate busy file server workload on multiple file systems in Linux, (5) \textit{Remote Direct Memory Access (RDMA)}: an artifact to test bandwidth and latency on local InfiniBand node, (6) \flashnet: neural network approach for fostering storage research in ML-based per-IO latency prediction.

Over the past two years, we have practiced using \tool{} in various settings. These include a seminar class focusing on reproducibility with over 20 students, Chameleon Summer REPRO 2020 with seven graduate researchers, and FAST'23 Birds-of-a-Feather (BoF) with 15 expected registrants. The FAST'23 BoF specifically focused on the extensible and reproducible experiment pattern called \textit{"breakable pattern"} methodology described in this work.

Through these opportunities, we have learned several lessons to further improve artifact reproducibility and \tool's usability for users. First, artifacts should be self-contained, as we cannot assume all users have prior knowledge or familiarity with associated papers. Second, authors should manage users' expectations throughout the reproduction process, including expected time consumption, output, figures, and other relevant information. Third, workflows should be idempotent to reduce user error and confusion. Finally, artifacts should only present immediate and relevant information to avoid overwhelming the user.

Overall, we show that it is plausible to create reproducible and extensible experiment artifacts. We conclude with many interesting discussions to
explore in the future. All experiments for \tool{} are made public.
\section{The Bleak State of Reproducibility in Storage Research \label{sec:ssd-bleak-state}}

Over the past decade, the storage market has experienced tremendous growth due to the explosion of big data and data-driven applications, such as machine learning (ML) and artificial intelligence (AI).
Along with the market, research in storage systems continues to thrive in academia. Despite the success in research, the storage system research community has historically neglected the issue of reproducibility, leading us to question the extent of its current impact.

Since many studies on storage systems~\cite{Liu+2019-SOML, Gao+2019-ReSSD, Yan+2017-Tiny, He+2017-Unwritten, Elyasi+2017-Slacker,Li+2016-AGCR}, particularly on solid-state disks (SSDs), have relied heavily on SSD simulators such as SSDsim~\cite{Hu+2011-SSDsim}, SimpleSSD~\cite{Jung+2017-Simplessd, Gouk+18-Amber}, and NANDFlash~\cite{Jung+2012-NANDFlashsim}. This situation arises because the majority of SSDs and their controllers/firmware are heterogeneous and proprietary to manufacturers. More importantly, the use of SSD simulators allows researchers to model and simulate all SSD characteristics and behavior along with the complete storage stack on standard CPUs (\eg{} x86 or arm64) and DRAM, eliminating the need for special hardware. Hence, in this paper, we will use the research papers using SSD simulators as the case study.
  
Leveraging simulators to conducting research should have made reproducing SSD research's \textit{artifact} (\ie{} the combination of the environment configurations, experiment scripts, and the analysis of the results) much more manageable than other types of research requiring special hardware (\eg{} ML/AI research with Graphics Cards or distributed and network systems research with Infiniband Network Interface Cards).
However, the reality is far from what is expected.

Throughout the case study, we investigate 193 SSD research papers published from 2010 to 2020 that used an SSD simulator for execution and evaluation. 
We determine the papers' artifacts' reproducibility formally based on six criteria: (1) \textit{Public} --- whether the artifact is publicly available and downloadable, (2) \textit{Compilation} --- whether the artifact can be compiled, (3) \textit{Execution} --- whether the compiled artifact can be run successfully, (4) \textit{Automation} --- whether the artifact is shipped with a script to automate the reproducing process, (5) \textit{Result} --- whether the results are consistent with the original researchers' result, and finally (6) \textit{Package} --- whether the artifact is packaged into an easily accessed container (\eg{} virtual machine or docker container).

\begin{table}[]
\centering
\resizebox{\linewidth}{!}{%
\begin{tabular}{@{}lccccccc@{}}
\toprule
\multicolumn{1}{p{1cm}}{\centering \textbf{Simulator\\ Used}} & \textbf{Total} & \textbf{Public} & \multicolumn{1}{p{1.25cm}}{\centering \textbf{Compil-\\ation}} & \multicolumn{1}{p{0.9cm}}{\centering \textbf{Exec-\\ution}} & \multicolumn{1}{p{0.9cm}}{\centering \textbf{Auto-\\mation}} & \textbf{Result} & \textbf{Package} \\ \midrule
SSDSim\cite{Hu+2011-SSDsim}    & 31 & 1 & 1 & 0 & 0 & 0 & 1 \\
MQsim\cite{Tavakkol+2018-MQsim}     & 2  & 1 & 1 & 1 & 0 & 0 & 0\\
SimpleSSD\cite{Jung+2017-Simplessd} & 35 & 4 & 3 & 1 & 1 & 1 & 1 \\
NANDFlash\cite{Jung+2012-NANDFlashsim} & 28 & 1 & 0 & 0 & 0 & 0 & 0 \\
gem5\cite{Binkert+2011-gem5}      & 3  & 1 & 0 & 0 & 0 & 0 & 0 \\
FEMU\cite{Li+2018-FEMU}      & 4  & 2 & 0 & 0 & 0 & 0 & 0 \\
FlashSim\cite{Kim+2009-Flashsim}  & 90 & 3 & 0 & 0 & 0 & 0 & 0 \\
\textbf{Total}     & \textbf{193}            & \textbf{13}          & \textbf{5}           & \textbf{2}         & \textbf{1}      & \textbf{1} & \textbf{2}       \\ \bottomrule
\end{tabular}%
}
\mycaption{tab-artifacts}{Artifacts Investigation}{
We investigate 193 artifacts published from 2010 to 2020 and found only 1 of them meets all the 6 criteria.
}
\end{table}

Surprisingly, as shown in Table~\ref{tab-artifacts}, only 13 artifacts (6.7\% of the total artifacts) are publicly available, 5 artifacts (2.6\% of the total artifacts) can be compiled, 2 artifacts (1\% of the total artifacts) are runnable, 1 artifact (0.5\% of the total artifacts) produces similar results compared to the original claimed results, and 2 artifacts are packaged (1\% of the total artifacts). This shocking revelation further motivate the development of this research project. These shocking results further urge and motivate us this paper.


\section{Chameleon Cloud and Trovi}

Chameleon \cite{Keahey+20-Chameleon} is an experimental testbed with the primary goal of supporting Computer Science research and education. The project, which has received full funding from the NSF FutureCloud \cite{NSF+14-NSFCloud}, has a budget of \$10 million and includes an impressive array of resources. These include 586 nodes (\ie bare metal servers), each with unique features such as GPU and FPGA nodes, as well as newly added storage nodes. The nodes are hosted in two locations, the University of Chicago and Texas Advanced Computing Center (TACC), and are connected using a high-speed 100Gbps network.

One of the key objectives of Chameleon is to provide its users with flexibility by leveraging the OpenStack ecosystem \cite{OpenStack} which is widely used in industry to manage huge numbers of nodes in clusters. This system is employed throughout Chameleon, enabling the management of OS images, instances, and user artifacts through object storage. By default, Chameleon offers a bare-metal experimental environment, where researchers have complete control over the machines, which allow low-level modifications such as operating system (OS) hacking and changing root privileges without any restrictions. In the event of a corrupted OS, the instance can be rebuilt in a single click, without losing any machine reservations. The web interface is intuitive and easy to use, providing users with full control over their instances, reservations, networks, OS images, and object storage which can be accessed at \cite{Chameleon-Cloud}.

For straightforward cases without the need for OS or other low-level modifications, Chameleon also grants simple virtual machines by utilizing Kernel-based Virtual Machine (KVM) which will spawn the instance faster due to their lightweight nature. Up until now, Chameleon provides seven configurations of KVM nodes which differ in their number of virtual CPUs, memory size and storage size. KVM nodes are similar to bare metal nodes except that there is no performance isolation. Thus, these types of nodes will be beneficial for users who do not need precise hardware measurements and profiling.

Chameleon also encourages its users to reproduce their research on the Chameleon platform by introducing the Jupyter Notebook Interface \cite{JupyterInterface} which is commonly used in data science and artificial intelligence fields. With Jupyter Notebook, users have the ability to attach their experiment scripts natively in Python programming language or shell scripts and execute these scripts in the Jupyter Notebook interface. Chameleon also integrates API access of Chameleon internal features, such as reservation and spawning instances using shell scripts and Python library \cite{Chameleon-PythonCHI}, by wrapping the OpenStack APIs to the Jupyter Interface.

Users will be able to set the experiment flows without any limitations. However, the most prominent feature of the Jupyter ecosystem lies in its replayability. Jupyter Notebook will save the output and variables obtained from the execution of the experiment code blocks. Hence, any important parts of the experiment can be re-run without re-executing the whole experiment and its preliminary steps, such as setting up the experiment environment and installing required softwares, subsequently minimizing the time to reproduce and modify the experiment. Indeed, we observed that the time needed to set up the experiment is profoundly longer than the runtime of the experiment itself. 

The reproducibility feature in Chameleon is taken further by the newly added Trovi platform \cite{Chameleon-Trovi} in the Chameleon ecosystem. Trovi facilitates sharing and uploading of experiment artifacts containing Jupyter Notebook files and scripts. This feature of Trovi platform allows other researchers to validate and reproduce the experiments carried out by other researchers. Trovi allows users to: (1) run the experiments on Chameleon nodes or (2) download the artifacts and run the experiments on other computational platforms. With almost 60 public experiments hosted on Trovi that are publicly accessible at \cite{Chameleon-Trovi}, the platform is an invaluable resource for researchers looking to collaborate and build upon each other's work.

\subsection{Chameleon Storage Infrastructure}

\label{sec:cc-storage-instances}

Chameleon is an innovative project that addresses the critical needs of storage research community. The introduction of new types of storage nodes is a testament to Chameleon's commitment to providing a better environment for storage research. To ensure that the storage nodes meet the needs of the researchers, the Chameleon team has ensured that the nodes are equipped with the latest technology such as solid-state drives (SSDs) and remote direct memory access (RDMA) over InfiniBand. In addition, Chameleon offers SSDs from various vendors to cater to a broader range of storage research needs.

The nodes in Chameleon typically come with a single solid-state drive or hard disk drive with ample space of more than 300GB. However, this is not sufficient for storage system researchers. The operating system's operations and other operations on users' data can cause additional performance impacts on the storage experiments if they are run altogether in a single storage device. To address this, storage nodes in Chameleon have been designed with one or more empty SSDs that support a redundant array of independent disks (RAID) by default. This gives researchers the flexibility and performance isolation they need to carry out their storage experiments with ease. The storage nodes have a capacity of more than 1TB, which is sufficient for storing experiment data and results. 

The Chameleon project is making significant strides in supporting the research community with its cutting-edge technology. Currently, the project has 30 bare-metal storage nodes that are capable of running storage experiments hosted at the University of Chicago and TACC that can be tweaked without any restrictions. To this end, the Chameleon project has announced plans to add more storage nodes in the near future. The additional nodes will be specifically designed to support more researchers and enable them to carry out and reproduce experiments within the Chameleon platform. The Chameleon project is committed to advancing research and reproducibility by supporting a growing number of researchers, as evidenced by their efforts to add more nodes and cutting-edge technologies. Hopefully, in the future, researchers will have more resources at their disposal to carry out their experiments, leading to faster and more reliable results.

Chameleon provides not only storage instances but also several mechanisms \cite{Chameleon-StorageManagement} that enable users to store and share their data effectively. Firstly, users can take a snapshot of their primary instance disk and create a bootable image containing all their data and configurations, which can be easily restored the next time they reserve servers on the Chameleon testbed. However, to mitigate longer boot times due to larger images, Chameleon provides binary objects storage, similar to enterprise-level storage options like Google Cloud Storage \cite{GoogleCloudStorage} and Amazon Web Services S3 \cite{AWS-S3}, with enhanced FUSE-based file system integration. This enables users to mount their storage anywhere in any instance. Additionally, to facilitate shared storage between bare-metal and virtual machines, Chameleon provides pre-allocated shared file systems utilizing OpenStack Manila \cite{OpenStack-Manila}, which allows users to detach file systems while maintaining data persistence. To further enhance the user experience, Chameleon also provides utility scripts to manage, mount, upload, and download data in the storage mechanisms mentioned above, helping researchers accelerate their development and empower reproducible research.

\subsection{Storage Research on Chameleon}

The Chameleon project is proven to be flexible for storage research by providing a wide range of storage nodes and disk vendors to researchers. These features have made it easier for researchers to evaluate their experiments, saving them from the hassle of acquiring multiple disks from different vendors, which can be an uphill task due to budgetary constraints or hardware scarcity in the market. As we all know, producing strong publications requires researchers to evaluate their works in multiple environments and vendors to demonstrate the efficacy of their research. With the availability of a range of storage nodes and disk vendors, storage researchers can now perform their experiments in different environments, leading to reliable research outcomes.

One of the works that benefits from the Chameleon platform is LinnOS \cite{MingzheHao+20-LINNOS}, published at OSDI 2020. They perform performance anomaly detection by leveraging latency prediction which becomes arduous over time because the device is getting faster. It is also well known that anomalies even in milliseconds granularity can cause monetary and productivity losses. In result, to adapt to this condition, they embrace the machine learning approach and build a model that can predict the speed, whether it is “fast” or “slow”, of every I/O that will be sent to an SSD device even at the millisecond granularity. \cite{MingzheHao+20-LINNOS} needs multiple enterprise-level SSD vendors to evaluate the efficacy of their model which fortunately is provided by Chameleon. The research team cited in \cite{MingzheHao+20-LINNOS} leveraged the Chameleon platform by replaying various storage traces on different devices to gather diverse performance profiles that were used to train their model. The support for heterogeneous SSDs provided by Chameleon played a significant role in enabling this type of research. In fact, LinnOS experiments can be accessed and reproduced at the Chameleon Trovi platform \cite{LinnOS-Trovi}.

There are also some other experiments that could have utilized the Chameleon platform. One of them is the IODA \cite{HuaichengLi+21-IODA} project published at SOSP 2021. IODA is an important research that aims to develop deterministic flash array design to provide predictable latencies that become a solution to severe service level objective (SLO) violations on a large scale. To fulfill the goal, the IODA team needs multiple SSDs for profiling tail-latency and evaluating software policy designs. On top of that, they also require experiments to run on top of bare flash arrays to show performance limitations of Linux software RAID running on an array of SSDs. For that, they bought a high-performant and expensive machine with at least more than 4 PCIe slots to attach 4 SSDs and other peripherals that need further additional flash arrays set up using Linux software RAID. Nowadays, Chameleon supports flash array research with 2 or more nodes that can accommodate 4 SSDs or more. These other projects, Harmonia \cite{Youngjae+11-Harmonia}, Gecko \cite{JiYong+13-Gecko}, FlashOnRails \cite{Dimitris+14-FlashOnRails}, Purity \cite{Colgrove+15-Purity}, SWAN \cite{Jaeho+19-SWAN}, RAIL \cite{Heiner+22-RAIL}, can also be adopted and reproduced in Chameleon storage infrastructure.

Chameleon also supports persistent memory (PM) servers equipped with 3TB Optane PM and 4-socket of 112 cores Intel Xeon processors. This enables experiments using PM such as PACTree \cite{Wook+21-PACTree} published in SOSP 21 in the Chameleon storage infrastructure. PACTree is an improved persistent key-value store on top of Intel Optane PM. PACTree team performed a thorough study of Optane PM performance characteristics to figure out performance and scalability of the real PM hardware. The main discoveries are twofold: (1) PM-optimized storage stacks may face a bottleneck at the PM bandwidth level and (2) the scalability of these stacks could be limited by the slow write latency of PM. In order to address these obstacles, they developed PACTree, a key-value store that utilizes Intel Optane PM. PACTree minimizes the usage of PM bandwidth and prevents slow PM latency from becoming a scalability limitation. This type of experiment requires a multi-core machine with a high number of cores along with Optane PM integration to evaluate the effectiveness and efficiency of PACTree. Not only PACTree \cite{Wook+21-PACTree}, other projects such as TIPS \cite{Madhava+21-TIPS} and TimeStone \cite{Madhava+20-TimeStone} can be run and reproduced on Chameleon infrastructure.

Recent works such as CrossFS \cite{Yujie+20-CrossFS}, FusionFS \cite{Zhang+22-FusionFS}, and KLOCs \cite{Sudarsun+21-KLOCs} focused on figuring out benefits, limitations and implications of near-storage computing and memory heterogeneity. CrossFS discovered the advantages of dividing file systems across userspace, the storage, and the OS to achieve better I/O performance and more sophisticated concurrency. FusionFS adopted and extended CrossFS design to implement CISC-style (Complex Instruction Set Computer) I/O operations that can naturally offload data for near-storage processing, reducing I/O operations such as data movement system calls and PCIe communication. On the other hand, KLOCs examined the advantages of memory heterogeneity using Optane memory. These works fortunately got the access to Optane PM \cite{IntelOptanePM} servers, however, emulating near-storage computing devices is still a rigorous work. The availability of Optane PM servers and near-storage computing devices (such as SmartSSD \cite{SmartSSD}) that natively supported in Chameleon would ease these types of research to: (1) conduct more detailed studies using enterprise-level devices and (2) explore new research areas in utilizing near-storage computing and storage heterogeneity.

\begin{figure*}[ht!]
\centerline{
 \includegraphics[width=\textwidth]{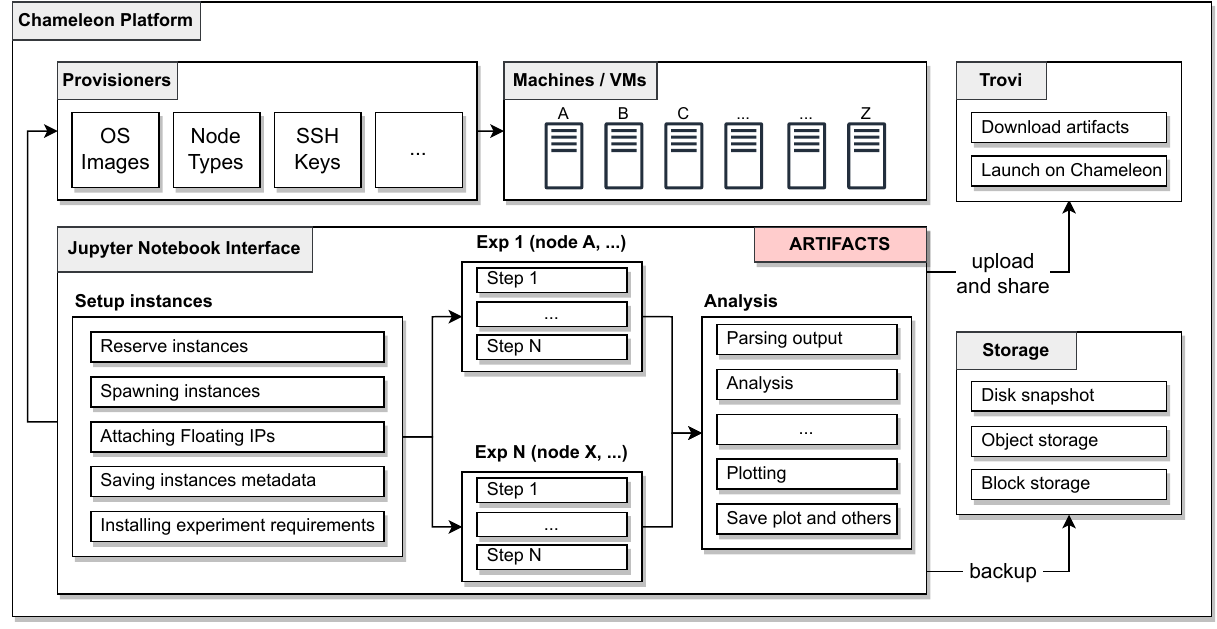}
}
\mycaption{exp-pattern}{"Breakable" experiment pattern}{Our opinionated, extensible and reproducible workflows by utilizing Jupyter Interface in Chameleon platform. Users can add their own personalized experiment on top of existing experiments by modifying existing notebook files or adding more notebook files.}
\label{fig:exp-pattern}
\end{figure*}

\section{Reproducible Storage Experiments on Chameleon}

\tool{} comprises a series of storage-related artifacts that can be easily reproduced and extended within the Chameleon infrastructure, which is publicly available through the Chameleon Trovi platform. \tool{} utilizes the Jupyter Notebook Interface and follows an opinionated and adaptable experiment pattern that we refer to as the \textbf{\textit{"breakable pattern"}} that is illustrated in Figure \ref{fig:exp-pattern}. Rather than relying on a single, unwieldy experiment script, we break the process into several manageable milestones.

\mypara{Setup} In this step, we allocate appropriate resources to serve as experiment containers in Chameleon, including the number of nodes, node types, network switches, router, and floating IPs. We then install or compile any necessary software, such as drivers or utilities, from various package managers and other source version control systems. Typically, the provisioning of experiment instances will take some time to complete, as we must spawn bare-metal instances, boot the operating system, and perform OS initialization tasks such as creating server users based on SSH keys and attaching disks to the server.

\mypara{Experimentation} Here, we can run the experiment code on the Chameleon nodes by executing scripts through an SSH connection based on the floating IPs we obtained in the setup stage. Users have the freedom to run the experiment code multiple times and adjust experiment configurations as desired, without needing to repeat the setup process. All experiment output generated by the experiment code is saved in files for later analysis.

\mypara{Analysis} This stage involves parsing and extracting crucial data from the output files generated during the experimentation stage. After identifying the key information, we encourage users to plot their results and save them in files, so that other users can view the anticipated outcomes when reproducing the experiments.

\mypara{Packaging} To enforce reproducibility, we recommend that researchers keep all experiment-related files, including experiment scripts, output files, plots, and any other generated materials, as part of their artifacts. These files can be useful to other researchers who wish to reproduce plots and view the analysis without running any experiments.

\mypara{Sharing} Finally, we motivate researchers to share their artifacts online. Chameleon provides the Trovi platform to facilitate researchers in uploading their artifacts and sharing them publicly, allowing other users to reproduce their artifacts inside Chameleon infrastructure. Native integration of Jupyter Notebook Interface and Chameleon platforms provides better user experience as other users only need to run whole artifacts without worrying about hardware reservation and software requirements. Even if other users do not want to reproduce the published artifacts in the Chameleon platform, Chameleon offers flexibility by allowing users to download the artifacts in a single compressed file. This compressed file will contain all the notebook files, experiment scripts, plots and generated files, if any. Chameleon also offers API and scripts that wrap OpenStack ecosystem utilities to perform OS snapshots and storing objects. This will speed up the experimental setup when reproducing the artifacts.

\mypara{Modification} Other users are encouraged and invited to create new artifacts by easily copy-pasting the setup, experiment and analysis code from their peers' artifacts. This extensible pattern will be helpful, especially in the context of education as it allows lecturers to give the base example that can be modified and extended freely, enhancing the students’ knowledge and creativity.

The most prominent benefit of adopting and integrating this pattern inside the Jupyter Interface is that users can extend and control the artifacts in any granularity. Users can easily break their milestones into several cells in a single Jupyter notebook file (coarse granularity) or in several Jupyter notebook files (fine granularity). None of them are better than the other. We use the former if we want to keep the variables and output that can be reused later in other cells inside the same notebook. The latter is used if users prefer more isolation such as running whole different experiments because each notebook file natively does not share variables and output to each other. Both granularities complement each other that can be seen in \ref{fig:exp-pattern}. In a single notebook file, such as provisioning instances, we split the big milestone into several small steps such as reserving instances, spawning instances and attaching the IP addresses to the spawned instances. However, when running multiple different sets of experiments, it is better to separate them into their own notebook file to provide better context and readability to other users.

In this work, we also focus on storage research reproducibility. As mentioned earlier in \ref{sec:cc-storage-instances}, Chameleon provides storage nodes to be able to run storage-related experiments by enabling performance isolation to provide better performance and noise-free outputs while running the experiment. Well known and battle-tested benchmarks in storage fields such as FIO \cite{Axboe-FIO} for storage device benchmarking and FileBench \cite{Filebench-FileBench} for file system benchmarking are used in several artifacts in this paper. Our main artifacts in this work are required to be run inside storage nodes. However, we commonly encounter situations where the storage nodes are fully reserved due to their high demand. The alternative experiment method is as follows: instead of running on an empty disk that is offered inside storage nodes, we run the artifacts on top of existing storage in any available nodes. This existing storage is not empty and already includes the OS, leading to reproducible experiment with the trade-off of having no performance isolation.
 
\subsection{Artifact 1: Simple FIO Benchmark}
\label{sec:exp:1-simple-fio-benchmark}
In this arifact, we run the FIO Benchmark \cite{Axboe-FIO} on a RAM disk (mounted using \textit{tmpfs}) and a SSD device. With the FIO Benchmark, the artifact is aiming to benchmark and compare the read throughput of the RAM disk and SSD. This artifact can be reproduced on virtually any machines on Chameleon Cloud as it does not require any special hardware.

\begin{figure}[!h]
\centerline{
 \includegraphics[width=\linewidth]{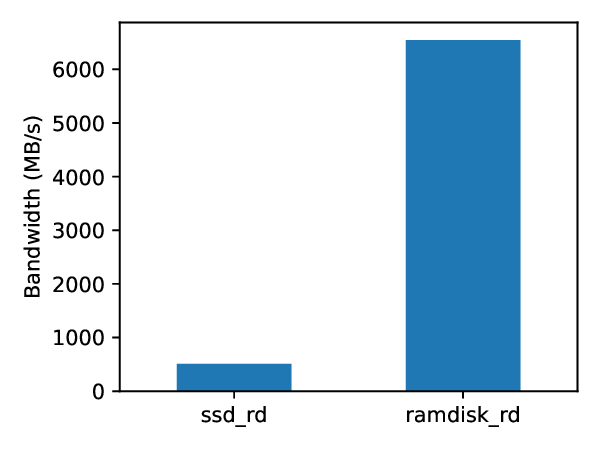}
}
\mycaption{fig-exp1}{Simple FIO Benchmark}{
Generated plot from analysis notebook file comparing the read throughput among single SSD and RAM disk.
}
\end{figure}

Following the breakable experiment patter, the artifact is separated into three notebooks: (1) environment setup (\eg{} provision servers, install dependencies, \etc{}), (2) run experiment based on a FIO benchmark configuration file which can be easily modified to align with users' interests (\ie{} specifying the block size for each read operation, the maximum read operations in queue, \etc{}), (3) analysis results by fetching and parsing the bandwidth results a ynd plot them into a bar chart shown in Figure \ref{fig-exp1}. The artifact can be accessed at \cite{Roy-SimpleFIOBenchmark}.

\subsection{Artifact 2: Simple File System Benchmark}
\label{sec:exp:2-simple-fs-bench}

In this simple experiment, we run multiple FileBench \cite{Filebench-FileBench} workloads for benchmarking default file system in Linux, \textit{ext4}. This experiment can be reproduced in any nodes in Chameleon or even in  personal computer or laptop. This experiment code simulates file system operations such as copying files, creating files, creating and removing directories, and performing heavy network file system operations. This  artifact is accesible at \cite{Ray+23-SimpleFilesystemBenchmarkTrovi}.

\begin{figure}[!ht]
\centerline{
 \includegraphics[width=\linewidth]{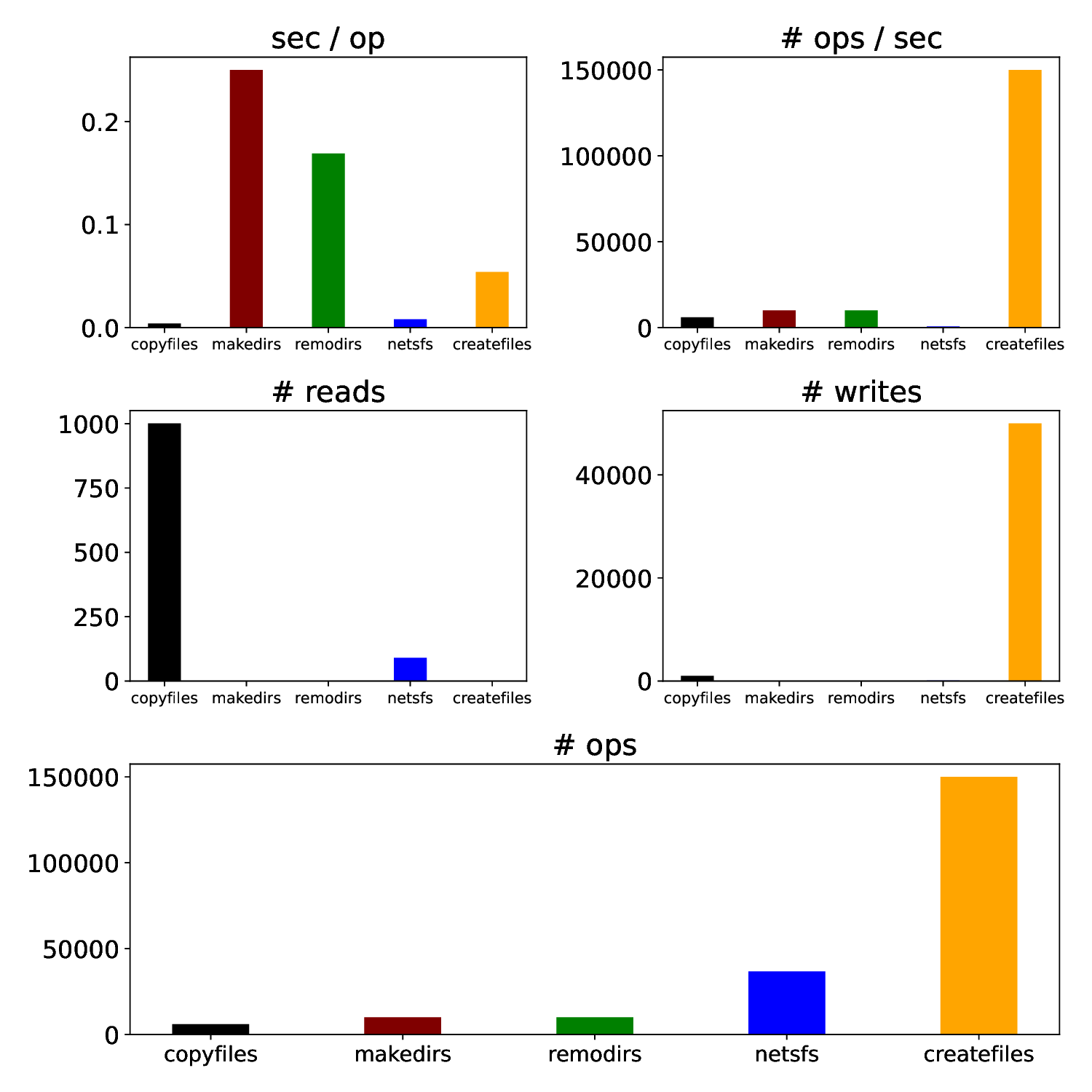}
}
\mycaption{simple-fs-benchmark}{Simple file system benchmark}{Generated plot from analysis notebook file showing performance of ext4 file system in different types of workloads.}
\label{fig:exp-2-simple-filesystem-benchmark}
\end{figure}

The artifact also breaks into several parts such as (1) setup experiment container, (2) run experiment code, and (3) analyze experiment outputs and generate plots. The overall execution time of this experiment should be done in less than 30 minutes including provisioning of bare-metal instances to execute the file system workloads. After running the experiment code, FileBench will produce the output to standard output stream containing important variables such as number of:  (1) seconds taken per operation, (2) operations per second, (3) operations, (4) reads, and (5) writes. The output of this experiment is shown in Figure \ref{fig:exp-2-simple-filesystem-benchmark} that is generated directly inside the analysis notebook file.

\subsection{Artifact 3: FIO Benchmark}

\begin{figure}[!h]
\centerline{
 \includegraphics[width=\linewidth]{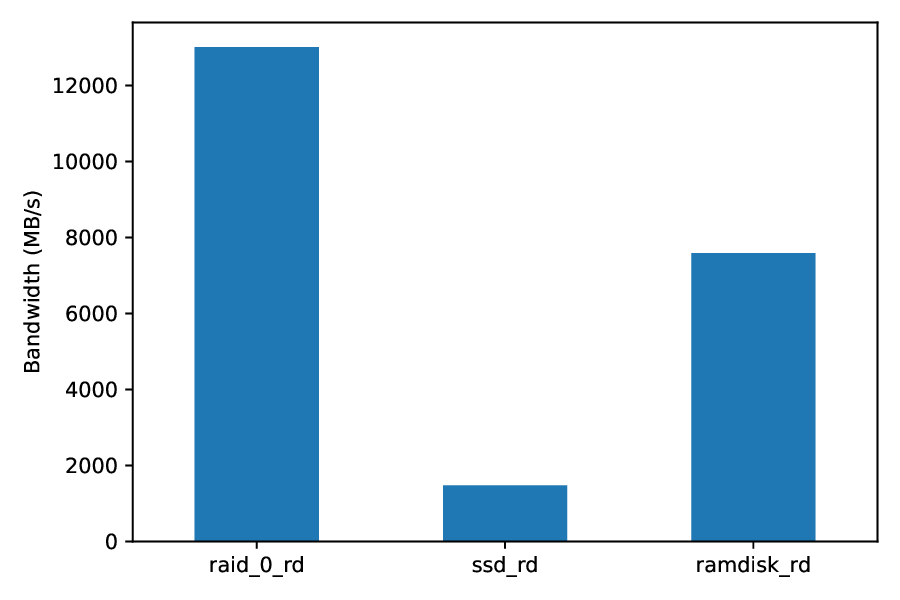}
}
\mycaption{fig-exp3}{FIO Benchmark}{
Generated plot from analysis notebook file comparing the read throughput among RAID-0 SSD array, single SSD, and RAM disk.
}
\end{figure}

This artifact further extends the Simple FIO Benchmark to benchmark the read throughput on a RAID-0 SSD array in addition to the RAM disk and SSD. As this artifact requires multiple SSDs, the artifact can only run on the storage node in the Chameleon Cloud. An example analysis figure are shown in Figure \ref{fig-exp3}. The artifact can be accessed at \cite{Roy-FIOBenchmark}.

\subsection{Artifact 4: File Systems Benchmark}

\begin{figure}[!hb]
\centerline{
 \includegraphics[width=\linewidth]{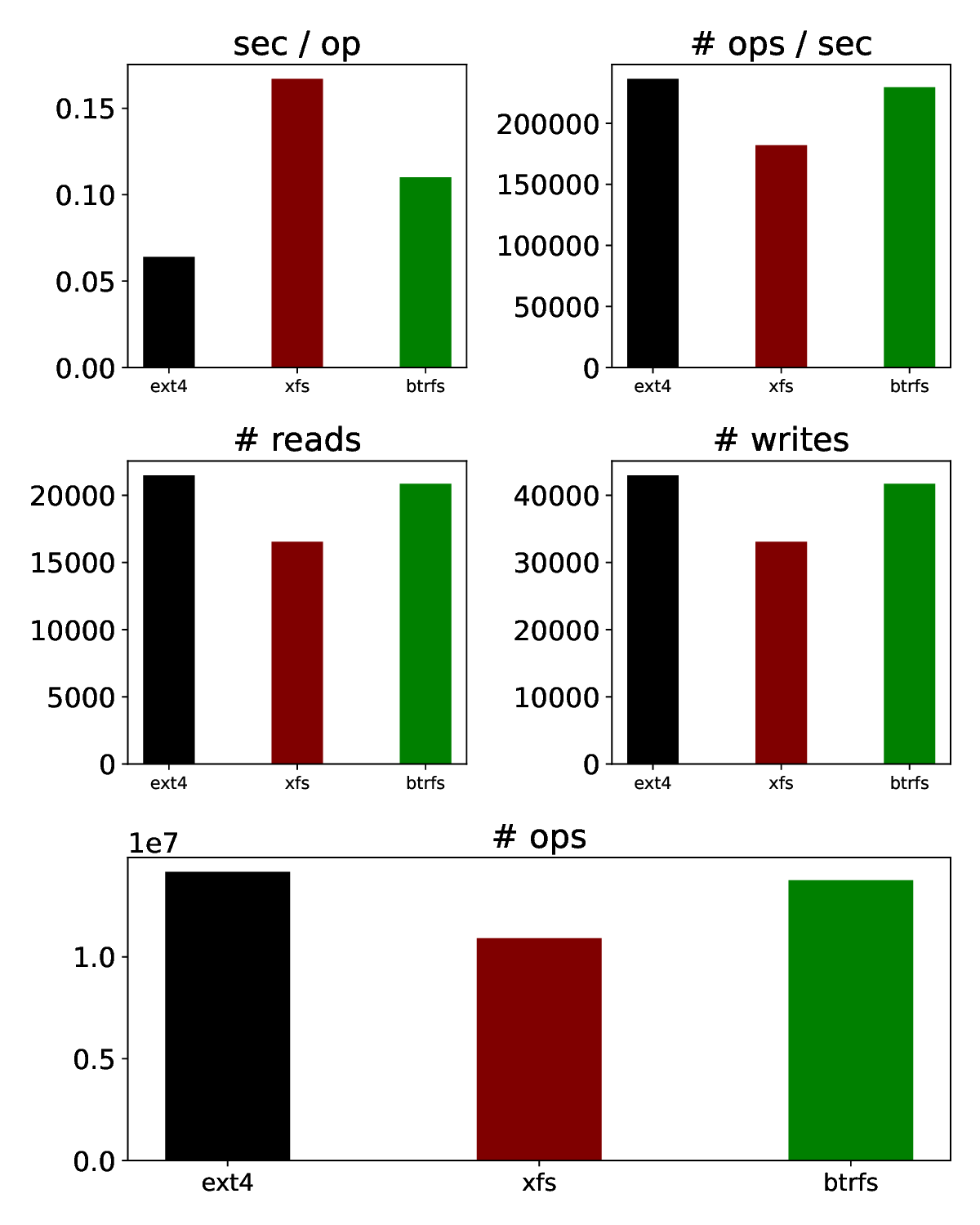}
}
\mycaption{filesystems-benchmark}{File systems benchmark}{Generated plot  showing performance of ext4, btrfs, and XFS file system in different types of workloads.}
\label{fig:exp-4-filesystems-benchmark}
\end{figure}

This is an enhanced and extended version of simple file system benchmark. Instead of running only on \textit{ext4} file system, we will try to run the file system workloads on other well-known Linux file systems such as \textit{Btrfs} and \textit{XFS}. This artifacts requires a storage node in Chameleon to utilize its additional empty SSD. The steps in this artifact are quite similar to Section~\ref{sec:exp:2-simple-fs-bench} with additional single step to prepare partitions to contain three file systems mentioned above. To create partition, we will format the empty disk and disable RAID partition, if any, because we do not need RAID features in this artifact. Then, we will create the three file systems inside that empty disk with the same configuration and enable minimum alignment as specified by the disk topology information. After completing the execution of the experiments code, the generated plot should be similar to Figure~\ref{fig:exp-4-filesystems-benchmark}. The variables that we used are originated from FileBench software as defined in Section~ \ref{sec:exp:2-simple-fs-bench}. This artifact is also published in Chameleon Trovi platform and can be accessed at~\cite{Ray+23-FilesystemsBenchmarkTrovi}.

\subsection{Artifact 5: Remote Direct Memory Access (RDMA)}

This simple artifact is intended to demonstrate capability of running RDMA benchmark on a local Chameleon node. RDMA is a mechanism to establish access to memory of two or more computers. InfiniBand is a popular implementation of RDMA that provides fast communication with very high throughput and very low latency and widely used in high performance computing (HPC) clusters. Fortunately, Chameleon provides support for utilizing InfiniBand in their platform by providing users with Mellanox~\cite{NvidiaMellanox} network device. Thus, specifically in this artifact, we reserve a node which has a Mellanox network device attached to it.


\begin{figure}[!h]
\centerline{
 \includegraphics[width=\linewidth]{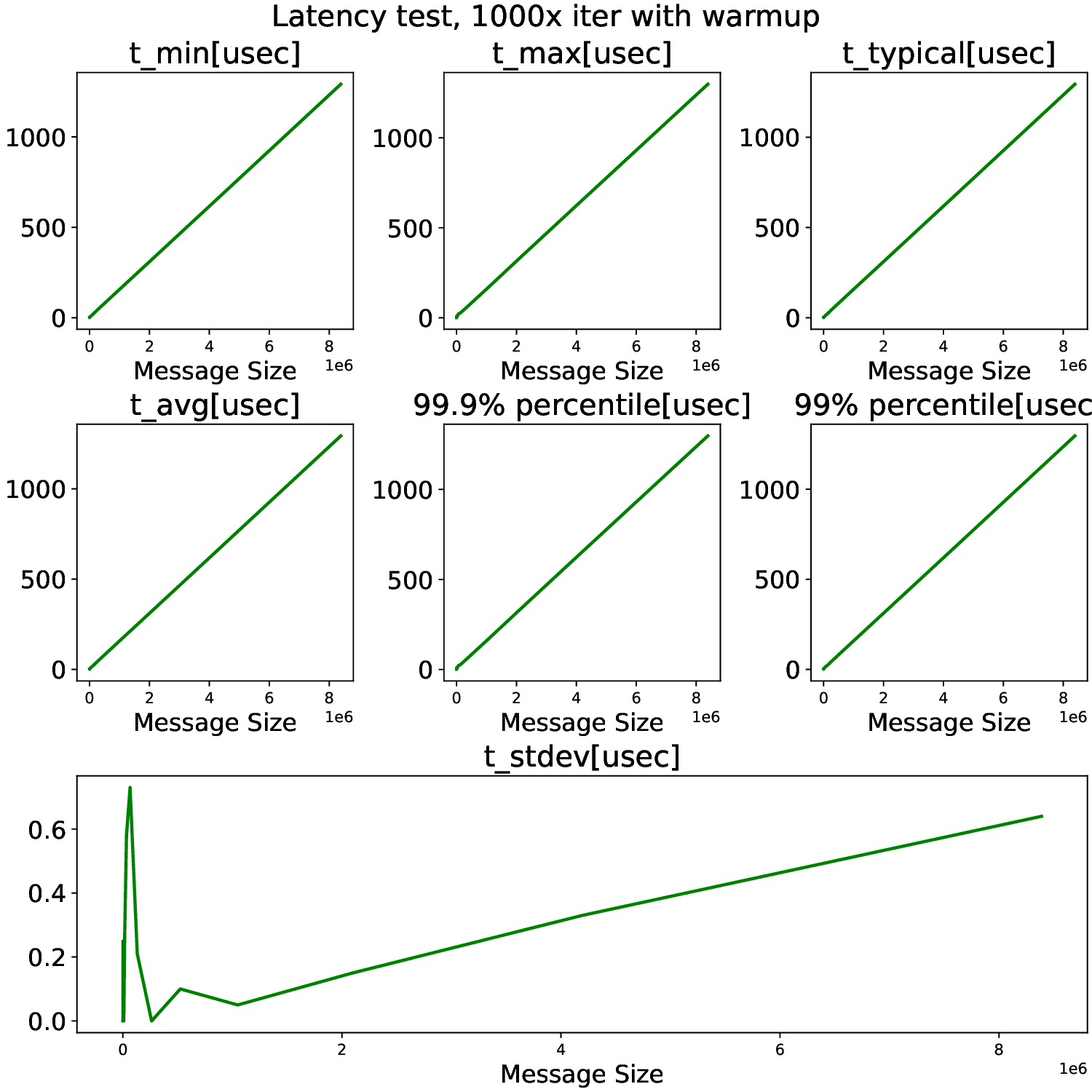}
}
\mycaption{rdma-latency-benchmark}{RDMA latency benchmark}{Produced plot from this artifact after running latency benchmark on a local InfiniBand node.}
\label{fig:exp-5-rdma-latency}
\end{figure}

The Mellanox driver itself comes with several numbers of benchmarking tools. In this artifact, users can measure the bandwidth of local InfiniBand network performance using a program called \textbf{ib\_send\_bw}. We also provide a script to run another benchmark to calculate latency using \textbf{ib\_send\_lat}. After completing this experiment code, users should see generated plots, one of them looking similar to Figure~\ref{fig:exp-5-rdma-latency}.

This artifact is publicly accessible at~\cite{Ray+23-RDMATrovi} and can hopefully be used as a stepping stone for other researchers to utilize and familiarize themselves with InfiniBand communication standards. One possible expansion of this artifact is the ability to run benchmarks in multiple InfiniBand nodes simultaneously to mimic production HPC clusters.

\subsection{Artifact 6: \flashnet, an ML for Storage Testbed}

The present study delves into the ongoing research called \flashnet, which employs a neural network (NN) model to perform per-I/O admission control. The primary objective of this research is to provide a benchmark for stimulating further investigation in the area of ML-based per-I/O latency prediction in storage research. The study is an extended version of \textsc{IONet} \cite{Daniar+21-IONet}. To train the model, the authors utilized Tencent I/O traces \cite{SNIA-IOTRACES}, which were replayed on their NVMe devices. Furthermore these traces is already preprocessed prior to their use for training the neural network model.

To carry out this research, the authors employed TensorFlow \cite{TensorFlow}, a neural network framework developed by Google. Though it is recommended to use GPU nodes for faster computation, the neural network model used in this study is relatively small and can be trained on CPU nodes. Therefore, users have the freedom to select their preferred node by modifying the experiment script. However, to ensure reproducibility, we opted to use CPU nodes since they are capable of executing the experiment adequately.

\begin{figure}[!h]
\centerline{
 \includegraphics[width=\linewidth]{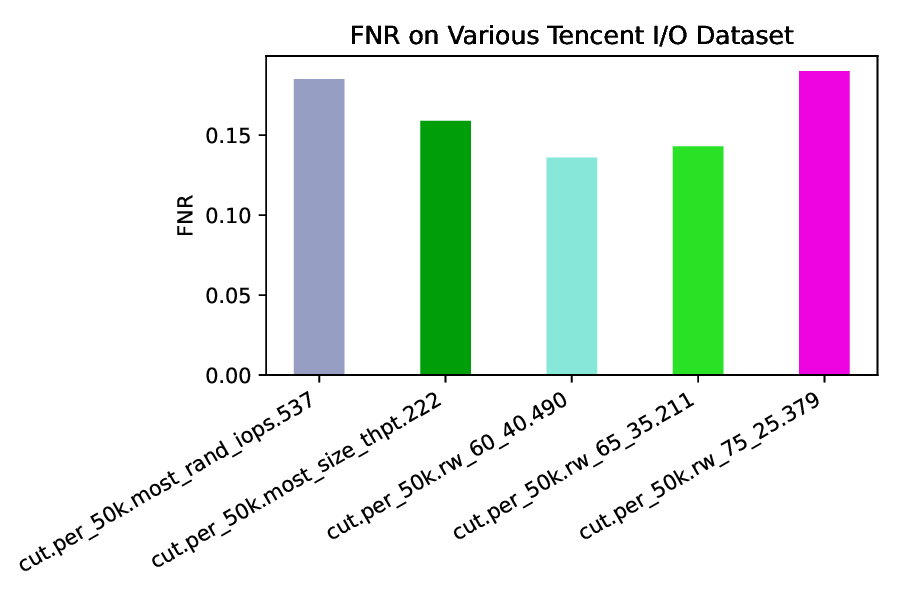}
}
  \mycaption{flashnet}{\textsc{FlashNet} performance}{False negative rate (FNR) is a metric that is used in \textsc{FlashNet} experiment for evaluating performance of the neural network model.}
\label{fig:exp-6-flashnet}
\end{figure}

The experiment is straightforward similar to usual neural network training pipeline and available at \cite{Ray+23-FlashnetTrovi}. First, we will label the data to denote whether the specific I/O  is "slow" or "fast" according to their trace. Then, we will extract feature from preprocessed data. Next, we will train the model using TensorFlow framework. Last, we will evaluate the model performance with the test data. After all steps are done, users should see the generated plot similar to Figure \ref{fig:exp-6-flashnet}.

\section{Lessons Learned}
Over the past two years, we have applied the methodology described above in a variety of scenarios beyond the \tool{}. These include classrooms with more than 20 students, the Chameleon Summer Repro 2020 event with 7 researchers, and the FAST'23 Birds-of-a-Feather event \cite{Fast23-BoF} (with an expected registration of over 15 attendees) from various academic levels (i.e. undergraduate, master's, and doctoral students). Alongside our use of the \tool{}, we provide the following lessons for leveraging \tool{} experiment pattern to package users' artifacts.

\mypara{Self-contain}
The artifact should be self-contained because we cannot assume that all users are familiar with the technical details of the original paper. Therefore, it is crucial to include all the necessary background, methodology, and preconditions (\eg{} access to resources, data sets, or recommended prior knowledge) in the artifact itself, rather than relying solely on the code.

\mypara{Managed Expectation}
Users will need to have a clear understanding of what to expect when reproducing this artifact. Providing the answers to the following questions int the artifact can help the users manage their expectations: 
\begin{itemize2}
\item How long is the artifact setup phase (\ie{} hardware or software configuration) expected to take? 
\item Does the artifact setup typically execute smoothly or are there some expected troubles? If so, are there ways to avoid them? 
\item How long is the execution of the artifact itself going to take? If the time of the artifact run is variable on the inputs, consider at least informing the user at execution time of the expected time to completion for the workflow step.
\item Are there any common error messages that users might run into?
\item Can you provide a visual representation of the anticipated outcome?
\end{itemize2}

\mypara{Building idempotent workflow}
It is common to re-run code cells in a notebook, and making sure that the cells are idempotent, if possible, can help reduce user errors and confusion for both the authors and users of the artifact. To this end, we provide a list of recommended procedures to maintain an idempotent workflow:
\begin{itemize2}
\item Avoid re-creating device reservations or re-initializing instances on the servers.
\item Avoid reinstalling packages unnecessarily. Most package managers should handle this automatically by default.
\item Avoid leaving any state (\eg{} global environment variables, file modifications, \etc{}) that may affect a cell's subsequent execution. For example, if you run a script on the server in cell A, and then modify it and re-run it in cell B, any user who runs cell A again will be executing the script with its modified contents, which can be confusing.
\end{itemize2}

\mypara{Handling errors gracefully}
Researchers usually expect some errors while building and running the experiments. To handle them, they need to utilize programming language features such as exception and error number to mitigate errors from libraries or softwares that they used. Furthermore, this error handling will also provide more contexts for users to help them independently debug their experiment execution.

\mypara{Avoid noise}
As the saying goes, ``Brevity is the soul of wit.'' If artifacts' authors include too much information, users may not see the big picture of the artifact. Therefore, we should avoid including information that is available elsewhere, in which case providing a link is best. Similarly, lengthy explanations of things that are not immediately relevant to the purpose of the artifact should also be avoided. Structuring the information in a way that highlights high-level concepts while allowing users to dig deeper for details is a bonus. Last, we also encourage researchers to sanitize their script execution output because sometimes the output is too verbose making it harder to pinpoint the problems in their artifacts.

\section{Conclusions and Discussions}

We are pleased to introduce \tool{}, a simple and practical approach to reproducibility research that addresses the pressing need for reliable and transparent storage-related experiment artifacts. As far as we are aware, \tool{} is the first reproducibility research that accommodates the Chameleon platform, empowering researchers to construct reproducible storage-related experiment artifacts.

Our research showcases the extensible nature of reproducible research without the need for cumbersome configuration of experiment containers, leveraging the Chameleon infrastructure. Additionally, we introduce a \textit{"breakable pattern"} that is possible thanks to the composability and flexibility of the Jupyter Notebook Interface. We are confident that this work will inspire further research in reproducibility across many fields of computer science.

With regard to extensibility and flexibility, the \textit{"breakable pattern"} we present in \tool{} is generalizable and adaptable to various types of user experiments. While our work is focused on Jupyter Notebook Interface and Chameleon platform, we acknowledge the potential for vendor lock-in. Thus, we hope to see more examples and guidelines that utilize this pattern in any platform or interface.

On the subject of reproducibility, \tool{} provides valuable advice and mechanisms for users to enhance the reproducibility of their research. However, creating pure idempotent and hermetic research artifacts is challenging and requires specific skills and platform support. These characteristics need to be enforced at the lowest possible level, but this can be difficult to realize on the hardware side due to the heterogeneous nature of hardware used by users seeking to reproduce experiments. Fortunately, we can enforce hermeticity and idempotence on the software side, at the operating system level.

In this regard, large-scale projects such as containerized builds using Docker \cite{Docker}, virtual machines, and pure hermetic operating systems like NixOS \cite{Nix-OS} can help solve this issue in the future. NixOS project consists of Nix \cite{Nix-PackageManager}, a purely functional package manager, and Nix functional programming language  \cite{Nix-ProgrammingLanguage}. The Nix package manager, inspired by functional programming paradigm, produces deterministic and hermetic builds by taking side effects into an account. This package manager relies heavily on a Nix functional programming language offers low-level OS and users configurations that can be transported anywhere, giving users a precise replica of the experiment software and OS used by the authors. Users can also easily review the source code, simplifying the process from their perspective.

In summary, \tool{} is a significant milestone in the advancement of reproducibility research, and we anticipate that it will stimulate further research in this area. We believe that the extensibility and flexibility of our \textit{"breakable pattern"} and the enforceable hermeticity and idempotence of the software side offer exciting prospects for researchers seeking to ensure the reliability, reproducibility, and transparency of their works. We are excited to continue contributing to the advancement of reproducible research in the future.

\singlespacing

{\small

   \bibliographystyle{plain}

   \bibliography{B/defs,B/all,B/personal,local.bib,B/confs}



}

\end{document}